\documentclass[twocolumn,showpacs]{revtex4}
\usepackage{graphicx}

\newcommand{\vecS}{\mbox{\boldmath $S$}}
\newcommand{\vecM}{\mbox{\boldmath $M$}}

\begin{document}
\title{Study of the Fully Frustrated Clock Model 
using the Wang-Landau Algorithm} 
\author{Tasrief Surungan}
\email{surungan@phys.metro-u.ac.jp}
\altaffiliation{On leave of absence from Department of Physics, Hasanuddin University, Makassar 90245, Indonesia}
\affiliation{Department of Physics, Tokyo Metropolitan University, Hachioji, Tokyo 192-0397, Japan}
\author{Yutaka Okabe}
\email{okabe@phys.metro-u.ac.jp}
\affiliation{Department of Physics, Tokyo Metropolitan University, Hachioji, Tokyo 192-0397, Japan}
\author{Yusuke Tomita}
\email{ytomita@issp.u-tokyo.ac.jp}
\affiliation{Institute for Solid State Physics, University of Tokyo,
Kashiwa 277-8581, Japan}

\date{Received 30 October 2003}

\begin{abstract}
Monte Carlo simulations using the newly proposed Wang-Landau algorithm 
together with the broad histogram relation are performed 
to study the antiferromagnetic six-state clock model 
on the triangular lattice, which is fully frustrated.  
We confirm the existence of the magnetic ordering belonging 
to the Kosterlitz-Thouless (KT) type phase transition 
followed by the chiral ordering which occurs at slightly higher temperature. 
We also observe the lower temperature phase transition of KT type 
due to the discrete symmetry of the clock model. 
By using finite-size scaling analysis, 
the higher KT temperature $T_2$ and the chiral critical temperature 
$T_c$ are respectively estimated as $T_2=0.5154(8)$ and $T_c=0.5194(4)$. 
The results are in favor of the double transition scenario.  
The lower KT temperature is estimated as $T_1=0.496(2)$.  
Two decay exponents of KT transitions corresponding to 
higher and lower temperatures are respectively estimated as 
$\eta_2=0.25(1)$ and $\eta_1=0.13(1)$, which suggests that 
the exponents associated with the KT transitions are universal 
even for the frustrated model. 
\end{abstract}

\pacs{05.70.Jk, 75.10.Hk, 75.40.Mg, 64.60.Fr}

\maketitle

\section{Introduction}
Frustration is one of the interesting subjects in statistical physics, 
mainly  because it can induce additional symmetry and lead the system 
to display rich low-temperature structures. 
The so-called two-dimensional (2D) fully frustrated XY models have attracted 
an extensive investigation in the last two decades 
\cite{vil,kawa,tei,grana,rami,lee1,ols,benak2,luo,boub,ozeki,miya,land,van,grana2,xu,lee2}.  
Due to the frustration the systems possess additional discrete reflection 
symmetry $Z_2$, apart from the global spin rotation symmetry $U(1)$. 
The breakdown of these symmetries are the onset of two types of 
phase transitions, namely one corresponding to the magnetic transition 
of Kosterlitz-Thouless (KT) type \cite{KT,Kosterlitz} and 
the other to the chiral transition. 
Whether these transitions are decoupled or occur at the same temperature 
has long been a controversy  
\cite{miya,grana,lee1,lee2,benak2,luo,land,van,grana2,xu,boub,rami,ols}. 
Another debated issue is whether the universality class of 
the chiral ordering belongs to the Ising universality class or not 
\cite{lee1,lee2,rami,ols}.

The system has a corresponding physical realization on a planar arrays 
of coupled Josephson junctions in a transverse magnetic field 
\cite{res,wees,eik,marconi} and discotic liquid crystals \cite{hal}.
As a 2D frustrated XY system, two lattice systems are 
frequently studied numerically. The first one is the square lattice 
where the interactions can be a regular mixture of ferromagnetic (F) 
and antiferromagnetic (AF) couplings (Villain model) 
\cite{vil,kawa,tei,grana,rami,lee1,ols,benak2,luo,boub,ozeki,grana2}. 
The second one is the AF XY model on the triangular lattice 
\cite{kawa,grana,benak2,ozeki,miya,land,van,grana2,xu,lee2}.  

As for the 2D XY model, the effect of the $q$-fold symmetry-breaking fields 
is an interesting subject \cite{Jose}; that is essentially the same 
as treating the $q$-state clock model, where only the discrete values 
are allowed for the angle of the XY spins. 
The $U(1)$ symmetry of the XY model is replaced by the discrete 
$C_q$ symmetry in the $q$-state clock model. 
It was shown \cite{Jose} that the 2D $q$-state clock model has 
two phase transitions of KT type at $T_1$ and $T_2$ 
($T_1<T_2$) for $q>4$.  There is an intermediate XY-like phase 
between a low-temperature ordered phase ($T<T_1$) 
and a high-temperature disordered phase ($T>T_2$). 
It is quite interesting to investigate the effect of 
the $q$-fold symmetry-breaking fields in the case of the 
fully frustrated XY model.  Quite recently, Noh {\it et al.} \cite{noh}
studied the AF six-state clock model on the triangular lattice using 
the Metropolis Monte Carlo simulation because of the experimental 
relevance to CF$_3$Br monolayers physisorbed on graphite \cite{fass}. 
However, they did not pay attention to the lower temperature phase transition 
of KT type. 

It is to be noticed that the existing controversy involves very fine values. 
Most studies claiming single transition scenario still do not exclude 
the possibility of two very close critical temperatures. 
Meanwhile, the studies in favor of double transition scenario 
always found that two critical phase transitions occur 
at slightly different temperatures.  
Therefore, it is desirable to obtain precise numerical information. 
Recently, much progress has been made in the development of 
efficient algorithms of Monte Carlo simulation.  
Especially, several attempts have been proposed for 
the Monte Carlo algorithms to calculate the energy 
density of states (DOS) directly.  Examples are the multicanonical method 
\cite{berg91,Lee93}, the broad histogram method \cite{oliv}, 
the flat histogram method \cite{jswang1,jswang2}, and the 
Wang and Landau method \cite{wl}. 
All of these algorithms use the random walk in the energy space. 

In this paper we report our Monte Carlo study on the AF six-state clock model 
on the triangular lattice. 
The ground state (GS) of the AF six-state clock model on 
the triangular lattice has the same structure as the AF XY model; 
therefore this model is regarded as a commensurate discrete model 
for the fully frustrated XY model.  
On the other hand, 
the six-state clock model on the square lattice (Villain model) 
has different GS configurations since there exist extra degeneracies. 
The presence of such extra degeneracy may bring about another interest 
in the fully frustrated six-state clock model . However, we will not 
cover such possibility in the present study.  
The XY Villain and the eight-state clock Villain models are 
commensurate because they have the same GS configuration. 

For the Monte Carlo method, we employ the Wang-Landau algorithm \cite{wl}, 
and the energy DOS is refined by the use of the broad histogram relation 
\cite{Oliv98,Berg98}. 
The fact that the energy of the six-state clock model is 
represented by the multiple of $J/2$, where $J$ is the coupling constant, 
is another supporting factor 
for the study of the six-state clock model;  it is convenient 
to treat discrete energy in the Monte Carlo simulation of 
calculating the DOS directly.

The rest of the present paper is organized as follows: In the next section 
we define the model and briefly explain the simulation method. 
Details of the calculation and results will be presented in Sec. III. 
The last section is devoted to the concluding remarks.

\section{Model and simulation method}
\subsection{Model and order}

The XY spin model is written with the Hamiltonian 
\begin{equation}\label{ham}
  H = \sum_{\langle ij \rangle} J_{ij} \, \vecS_i \cdot \vecS_j 
    = \sum_{\langle ij \rangle} J_{ij} \cos(\theta_i-\theta_j),
\end{equation}
where $\langle ij \rangle$ denotes the summation over nearest neighbor 
interactions, $\vecS_i$ a unit planar spin vector occupying the $i$-th site, 
and $\theta_i$ the angle associated with the $i$-th spin. 
Here, we mainly study the six-state clock model; 
therefore the angle takes discrete values, $\theta_i = 2\pi p/6$ 
with $p=0,\cdots,5$. The frustration is conveyed by $J_{ij}$. 
For the Villain model on the square lattice this can be set 
by taking regular mixture of F and AF couplings. 
For the triangular lattice on the other hand, $J_{ij}$ are simply set 
to be uniform AF couplings, $J_{ij}=J>0$, so that the system becomes 
fully frustrated. 

\begin{figure}
\includegraphics[width=1.0\linewidth]{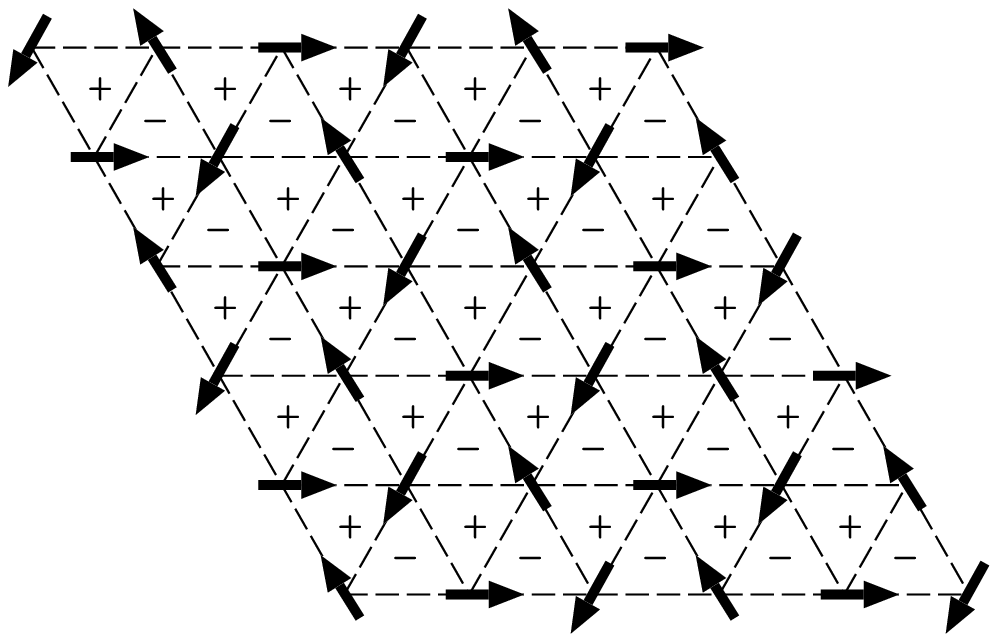}
\caption{ A ground state configuration of the AF six-state clock model 
on the triangular lattice of size $6\times 6$.  
Spins are represented by arrows.  
Sites belonging to the same sublattice have the same orientation of spins. 
The $+$ and $-$ signs indicate the handedness of the local chiralities. 
The ground state has 12-fold degeneracy.}
\label{gsconf}
\end{figure}

The Hamiltonian (\ref{ham}) is invariant under the symmetries of 
the global spin rotation $U(1)$ and the global spin reflection $Z_2$. 
The breaking of these symmetries is expected to cause two kinds of ordering, 
which respectively correspond to magnetic ordering and chiral ordering.  
The GS configuration is well known as $2\pi/3$-configuration, 
where two neighboring spins align in $2\pi/3$ difference in angle, 
which is shown in Fig.~\ref{gsconf}.  
We decompose the lattice into three interpenetrating sublattices for studying 
magnetic order.  A site in a triangle belongs to one of the sublattices, 
$A, B$ or $C$.  We assign the magnetic order parameter as 
\begin{equation}
\label{ordp1}
 m^2=\frac{3}{N^2}\left({M_A}^2+{M_B}^2+{M_C}^2 \right),  
\end{equation}
where $\vecM_A=\sum_{i \in A} \vecS_i$ is the magnetization of sublattice $A$, 
and $N$ is the number of spins; 
$\vecM_B$ and $\vecM_C$ follow the same definitions for the sublattices $B$ and $C$.

To discuss the global spin reflection $Z_2$, we deal with the chirality.  
The local chirality on the elementary triangle is defined as 
\begin{equation}
\label{chiral}
\kappa_i=\frac{2}{3\sqrt 3} \sum_{j,k \in {\triangle}}{[\vecS_j \times \vecS_k]}_z 
 = \frac{2}{3\sqrt 3} \sum_{j,k \in {\triangle}}\sin(\theta_k-\theta_j),
\end{equation}
where the $z$ component of the vector chirality is considered. 
The numerical factor in Eq.~(\ref{chiral}) is chosen such that 
the maximum of the absolute value is one.  
In the GS configuration depicted in Fig.~\ref{gsconf}, 
the local chirality takes a checkerboard pattern of 
the right-handed (positive) orientation and 
the left-handed (negative) orientation. 
Then, the staggered chirality 
\begin{equation}
\label{ordp2}
   \kappa = \frac{1}{2N} \sum_i (-1)^i \, \kappa_i
\end{equation}
becomes the order parameter 
for the $Z_2$ symmetry breaking transition. 

The GS configuration has $12$-fold degeneracy which is induced 
by the discrete global spin rotation symmetry $C_6$ with $6$-fold and 
by $Z_2$ symmetry with $2$-fold.  The number of this degeneracy is 
used as one of the check conditions in the calculation of energy DOS.

\subsection{Simulation method}

We use the Monte Carlo method to calculate the energy DOS directly 
to obtain precise numerical information.  
First, we briefly describe the Wang-Landau algorithm \cite{wl}. This algorithm is similar to the multicanonical method (entropic sampling) of Lee \cite{Lee93}, the broad histogram method \cite{oliv} and the flat histogram method \cite{jswang1,jswang2}; 
the idea is based on the observation that performing a random walk in energy space with a probability proportional to the reciprocal of the DOS, $1/g(E)$, will result in a flat histogram of energy distribution. 
The Wang-Landau method introduces a modification factor 
to accelerate the diffusion of the random walk in the early stage of the 
simulation. 
Since the DOS is not known at the beginning, it is simply set $g(E)=1$ for all energy $E$. The transition probability from energy $E_1$ to $E_2$ reads
\begin{equation}
 p(E_1 \rightarrow E_2)={\rm min} \left[\frac{g(E_1)}{g(E_2)},1 \right],
\label{trans}
\end{equation}
and the DOS $g(E)$ is iteratively updated as 
\begin{equation}
\label{mod}
 \ln g(E) \rightarrow \ln g(E) + \ln f_i 
\end{equation}
every time the state is visited. The modification factor $f_i$ is 
gradually reduced to unity by checking the `flatness' of the energy histogram; 
the histogram for all possible $E$ is not less than some value of 
the average histogram, say, 0.80. 

We also use the broad histogram relation for getting a refined DOS. 
In proposing the broad histogram method, 
Oliveira {\it et al.}~\cite{oliv} paid attention to 
the number of potential moves, or the number of 
the possible energy change, $N(S, E \to E')$, 
for a given state $S$.  The total number of moves is
\begin{equation}
 \sum_{\Delta E} N(S, E \to E + \Delta E) = N
\end{equation}
for a single spin flip process of the Ising model simulation. 
The energy DOS is related to the number of potential moves as
\begin{equation}
 g(E) \, \left< N (S, E \to E') \right>_E  
   = g(E') \, \left< N (S', E' \to E) \right>_{E'}, 
\label{BHR}
\end{equation}
where $\left< \cdots \right>_E$ denotes the microcanonical 
average with fixed $E$.  This relation is shown to be valid 
on general grounds \cite{Oliv98,Berg98}, and we call 
Eq.~(\ref{BHR}) the broad histogram relation. 
We measure the average of the potential move, 
$\left< N (S, E \to E') \right>_E$, and use this 
information for getting a better estimate of the energy DOS. 
It was stressed \cite{Oliv00,Lima00} that $N(S, E \to E')$ 
is a macroscopic quantity, which is the advantage of using 
the number of potential moves.  
%%%
We should also note that the broad histogram relation does not
depend on the particular dynamic rule one adopts, and 
the microcanonical averages of the potential moves can be 
obtained by any rule of Monte Carlo dynamics.  
%%%

In order to reduce calculation time for larger system sizes, 
we break simulation into several energy windows and perform random walk 
in each different range of energy. The resultant pieces of the DOS are 
joined together and used to produce the thermal average with 
the inverse temperature $\beta$ through the standard relation
\begin{equation}
\label{ave}
 \left< Q \right>_{\beta}=\frac{\int Q(E)g(E) e^{-\beta E}dE}{\int g(E)e^{-\beta E}dE}.
\end{equation}
Using the parallel machine, we perform the measurements of 
the physical quantity $Q$ up to $64 \times 10^5$ Monte Carlo steps.  
Also, we perform 10 independent runs for each system size 
in order to get better statistics and to evaluate statistical errors.

\section{Results}
\subsection{Energy DOS and specific heat}

\begin{figure}
\includegraphics[width=1.0\linewidth]{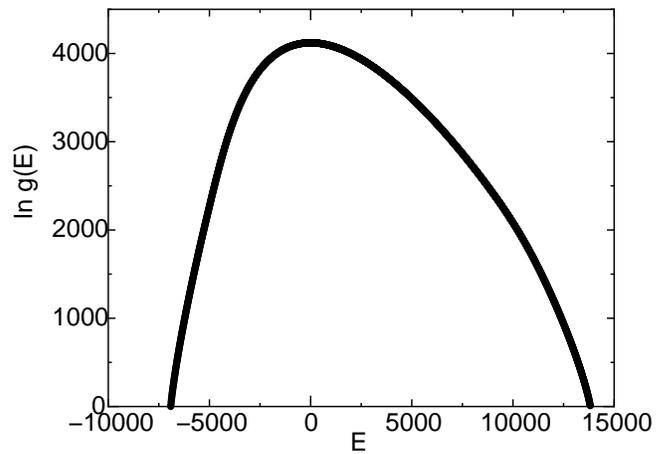}
\caption{Energy DOS of system size $L=48$. 
The energy is represented in units of $J/2$.}
\label{ds48} 
\end{figure}

Here we present the results for the AF six-state clock model on the triangular lattice.  
We have treated the system with the linear sizes $L$ = 24, 36, 48, 60, and 72. 
We apply the periodic boundary conditions. 
We normalize the DOS by using the condition $\sum_E g(E) = 6^N$, 
and the degeneracy in the GS energy, $g(E_{\rm GS})=12$, is checked 
in order to confirm the accuracy of the calculation. 
In Fig.~\ref{ds48}, we show the energy DOS of system size $L=48$ 
as an example.  Here, the energy is represented in units of $J/2$, 
and the GS energy is given by $-(3/2)NJ$. 

The energy-dependent data of quantity $Q(E)$ are used to calculate 
the thermal average $\langle Q \rangle$ by using Eq.~(\ref{ave}).  
We calculate the specific heat per spin through the relation 
\begin{equation}
 C(T)=\frac{1}{Nk_{\rm B}T^2}[\langle E^2 \rangle - \langle E \rangle^2], 
\end{equation}
where $k_{\rm B}$ is the Boltzmann constant.

\begin{figure}
\includegraphics[width=1.0\linewidth]{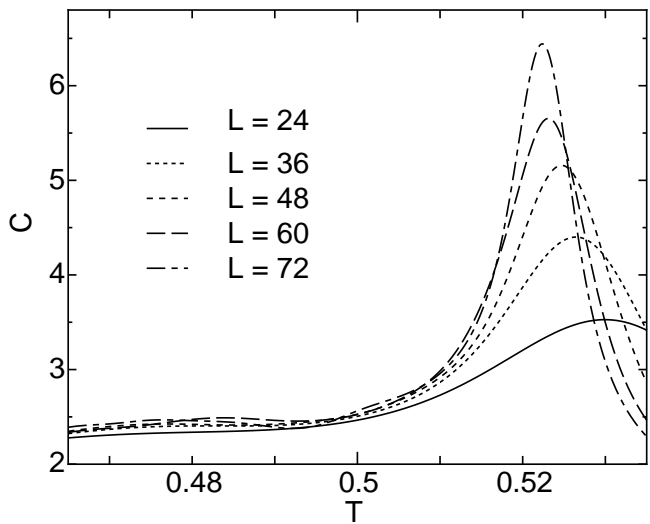}
\caption{Temperature dependence of specific heat 
for system size $L$ = 24, 36, 48, 60, and 72.}
\label{spht} 
\end{figure}

We show the temperature dependence of specific heat 
for different lattice sizes in Fig.~\ref{spht}.  
The divergent peak around $T \simeq 0.52$ in units of $J/k_{\rm B}$ 
gives a clear sign of the existence of second-order 
phase transition.  We also observe a hump 
on the lower temperature side around $T \simeq 0.48$, 
which may be related to the transition of KT type. 
However, we should study the magnetic and chiral orders 
for the detailed analysis of the phase transition. 

\subsection{Correlation ratio}
The critical behavior and the transition temperature can be investigated 
more precisely from the evaluation of the order parameter or 
its corresponding correlation function.  The magnetic and chirality 
correlation functions are defined as the following: 
\begin{eqnarray}\label{magchi1}
 G(r) &=& \langle \vecS_{i} \cdot \vecS_{i+r} \rangle, \\ \label{magchi2}
 \gamma(r) &=& \langle \kappa_{i} \, \kappa_{i+r}\rangle, 
\end{eqnarray}
where $r$ is the fixed distance between spins.  
Precisely, the distance $r$ is a vector, but we have used 
a simplified notation. 

Two of the present authors \cite{gpcc} showed that 
the ratio of the correlation functions with different distances 
is a useful estimator for the analysis of the second-order phase transition 
as well as for the KT transition, and this correlation ratio can be 
used for the generalization of the probability-changing 
cluster algorithm \cite{pcc}. 

At the critical point or on the critical line, 
the correlation function $g(r)$ for an infinite system 
decays as a power of $r$, 
\begin{equation}
  g(r) \sim r^{-(D-2+\eta)}, 
\label{g(r)}
\end{equation}
where $D$ is the spatial dimension and $\eta$ the decay exponent. 
For a finite system in the critical region, 
the correlation function depends on two length ratios,
\begin{equation}
  g(r,t,L) \sim r^{-(D-2+\eta)}h(r/L,L/\xi),
\label{corr_finite}
\end{equation}
where $\xi$ is the correlation length.
Then, the ratio of the correlation functions with different 
distances has a finite-size scaling (FSS) form with a single scaling variable,
\begin{equation}
  \frac{g(r,t,L)}{g(r',t,L)} = f(L/\xi),
\label{corr_ratio}
\end{equation}
if we fix two ratios, $r/L$ and $r/r'$.

In the present work, we set $r=L/2$ and $r'=L/4$ 
for two distances.  Thus, we evaluate the correlation ratios 
$G(L/2)/G(L/4)$ and $\gamma(L/2)/\gamma(L/4)$,
where $G$ and $\gamma$ are referred respectively to Eqs. (\ref{magchi1}) 
and (\ref{magchi2}).
It is important for two correlated spins to belong to the same sublattice; 
since the fixed distances are set as $r=L/2$ and $L/4$, we choose 
the system size as a multiple of $12$.  

\subsubsection{Kosterlitz-Thouless transitions}

\begin{figure}
\includegraphics[width=1.0\linewidth]{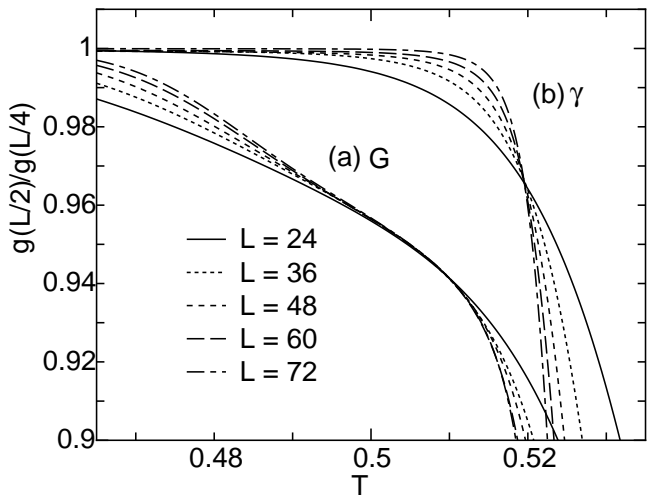}
\caption{Temperature dependence of ratios of 
the (a) magnetic and (b) chiral correlation functions.}
\label{corf} 
\end{figure}

We show the correlation ratios both for the (a) magnetic 
and (b) chiral correlations in Fig.~\ref{corf}. 
From the temperature dependence of the magnetic correlation ratio 
plotted in Fig.~\ref{corf}(a), 
we observe that the curves of different sizes 
merge in the intermediate temperature range ($T_1<T<T_2$), 
and spray out for the low-temperature and high-temperature ranges. 
This behavior is the same as that for the unfrustrated six-state 
clock model \cite{gpcc}, which suggests that there are 
two phase transitions of KT type at $T_1$ and $T_2$. 
The hump on the lower temperature side 
in the specific heat, Fig.~\ref{spht}, 
may correspond to the lower temperature KT transition at $T_1$.  
The higher temperature KT transition at $T_2$ is not obvious 
from the specific heat plot as it is veiled by the divergent peak 
due to the chiral transition. 

\begin{figure}
\includegraphics[width=1.0\linewidth]{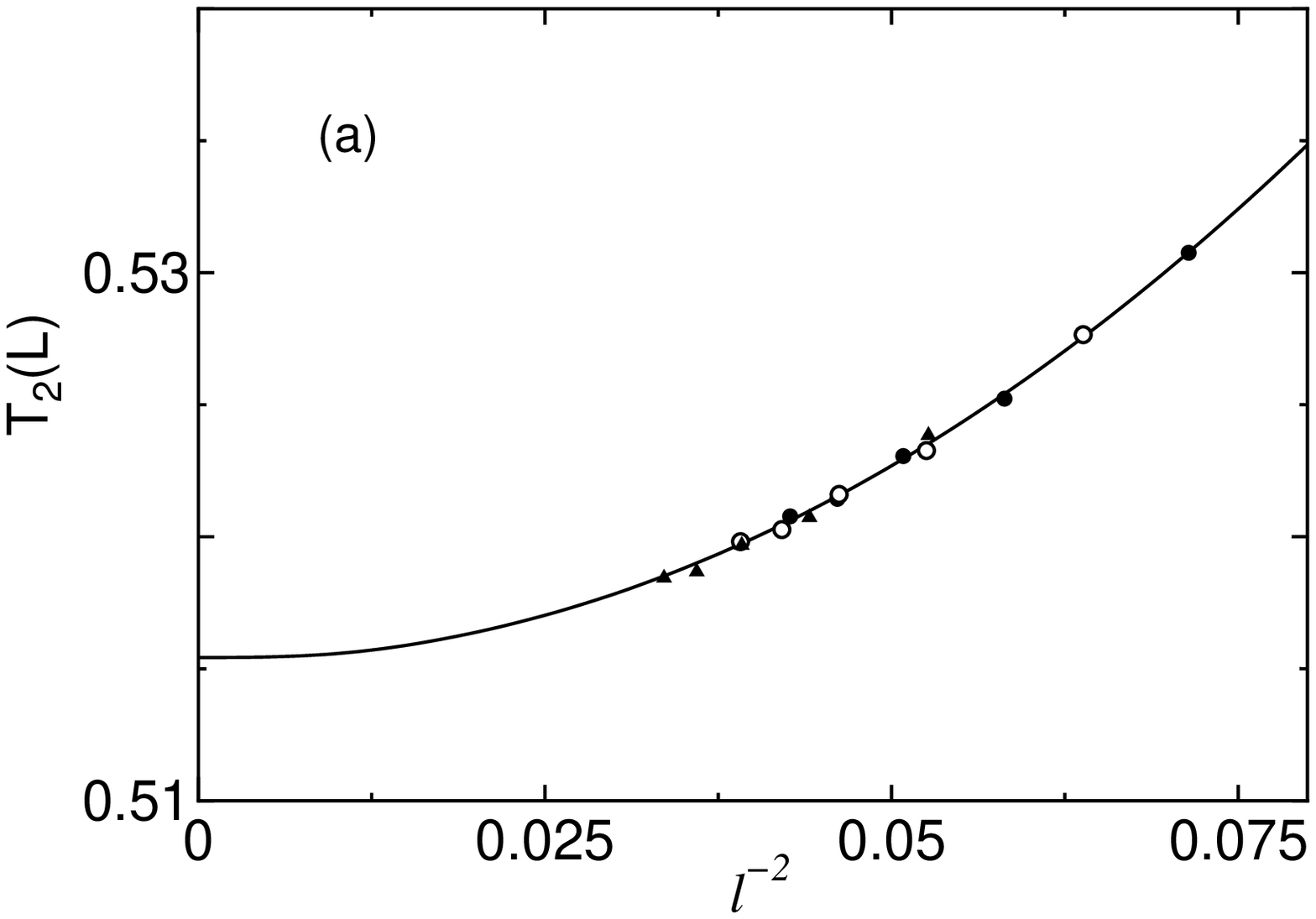}
\\
\vspace{4mm}
\includegraphics[width=1.0\linewidth]{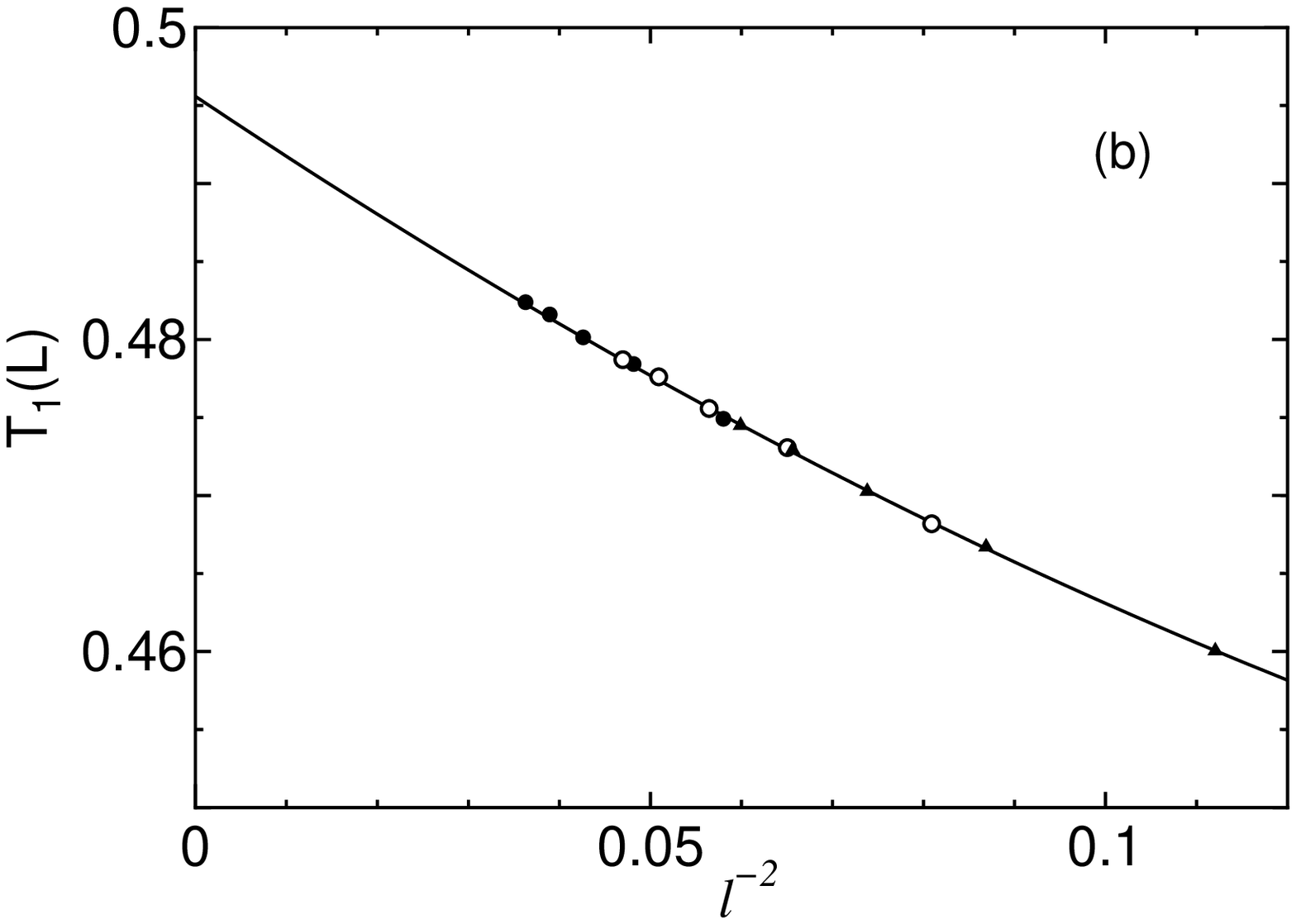}
\caption{Plot of (a) $T_2(L)$ and (b) $T_1(L)$ of the AF six-state 
clock model on the triangular lattice 
for $L$ = 24, 36, 48, 60, and 72, where $l = \ln bL$. 
The data for $R$=0.86, 0.88 and 0.90 are shown 
by different marks in (a), and those 
for $R$=0.99, 0.985 and 0.98 in (b).
}
\label{T12} 
\end{figure}

We can make a FSS analysis based on 
the KT form of the correlation length, 
$\xi \propto \exp(c/\sqrt{t})$,
where $t=|T-T_{\rm KT}|/T_{\rm KT}$.  
The $L$ dependence of $T_{\rm KT}(L)$ is given by 
\begin{equation}
 T_{\rm KT}(L) = T_{\rm KT} + \frac{c^2T_{\rm KT}}{(\ln bL)^2}.
\label{T_KT}
\end{equation}
Using the data of the magnetic correlation ratio 
$R=G(L/2)/G(L/4)$ for different sizes, 
we estimate two KT transition temperatures. 
We consider the size-dependent temperature that gives 
the constant $R$. 
In Fig.~\ref{T12}, we plot $T_{\rm KT}(L)$ as a function of $l^{-2}$ 
with $l=\ln bL$ for the best-fitted parameters in Eq.~(\ref{T_KT}). 
For a fitting function we have used a quadratic function in $l^{-2}$ 
to include correction terms. 
The value of $R$ has been set to be 0.86, 0.88 and 0.90 
for the determination of $T_2$, whereas $R$ has been set to 
be 0.99, 0.985 and 0.98 for $T_1$. 
The data with different $R$ are represented by different 
marks in Fig.~\ref{T12}, but they are collapsed 
on a single curve in this plot, which means that 
$b$ depends on $R$ in Eq.~(\ref{T_KT}) and the difference of 
$R$ can be absorbed in the $R$ dependence of $b$. 
We estimate the KT temperatures 
of the magnetic order using Eq.~(\ref{T_KT}) as 
$$
  T_2=0.5154(8) \quad {\rm and} \quad T_1=0.496(2), 
$$
where the numbers in the parentheses denote the uncertainty 
in the last digits.  
The estimate of $T_2$ is slightly lower than the estimate 
by Noh {\it et al.} \cite{noh}, $0.5175(3)$.  It is due to 
the fact that the moment ratio was used in Ref.~\cite{noh}, 
and the estimate of the KT temperature becomes higher 
because of large corrections to FSS \cite{gpcc}.  

\begin{figure}
\includegraphics[width=1.0\linewidth]{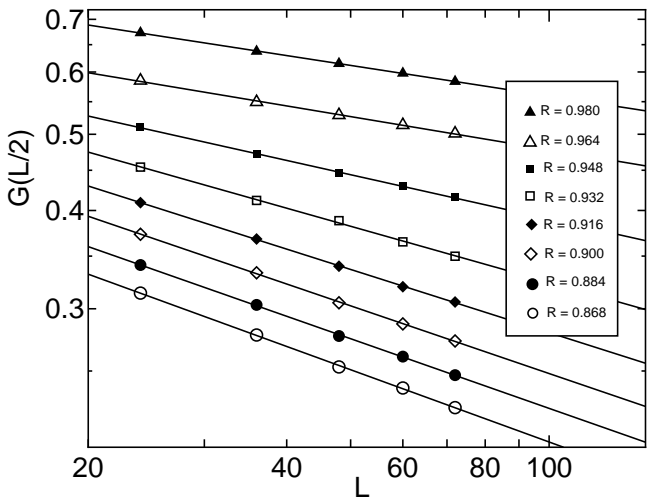}
\caption{Plot of correlation function $G(L/2)$ versus $L$. Here the slope of the best-fit straight line of each corresponding $R$ is the value of exponent $\eta$.}
\label{slope} 
\end{figure}

Next we consider the decay exponent $\eta$.  
We first look at the constant value of correlation ratio $R$
for different sizes and find the associate correlation function $G(L/2)$. 
We give attention to the power-law dependence of the correlation function 
on the system size, $G(L/2) \sim L^{-\eta}$, which can be 
seen from Eq.~(\ref{corr_finite}) and $D$ is set to be 2. 
We plot $G(L/2)$ versus $L$ for various $R$'s 
in logarithmic scale in Fig.~\ref{slope}. 
The value of $\eta$ is obtained as the slope of 
the best-fitted line for each constant of correlation ratio. 
The multiplicative logarithmic corrections for the KT transition 
\cite{Kosterlitz,Janke97} were shown to be small 
compared to statistical errors. 

\begin{figure}
\includegraphics[width=1.0\linewidth]{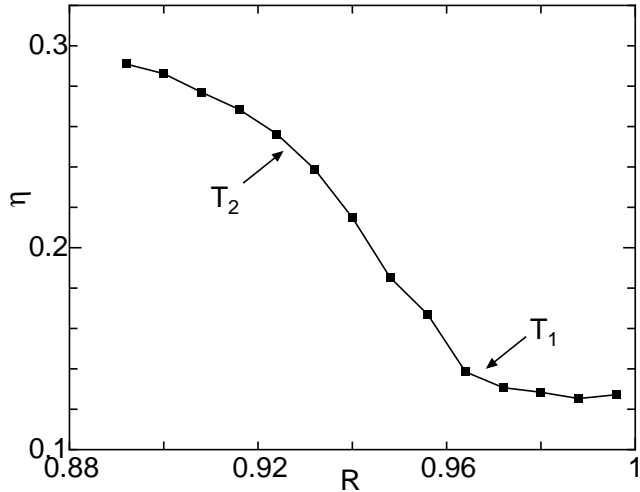}
\caption{Decay exponent $\eta$ of KT phase as a function of magnetic coefficient ratio $R$.  Line is just guide to the eyes.}
\label{eta} 
\end{figure}

We plot $\eta$ thus determined with respect to the fixed correlation 
ratio $R$ in Fig.~\ref{eta}.  In the KT phase, $R$ is 
directly related to the temperature. 
We should note that the exponent $\eta$ is meaningful only in the 
temperature range $T_1 \le T \le T_2$ on the fixed line.  
We show the values of $R$'s 
which give $T_1$ and $T_2$ by arrows in Fig.~\ref{eta}. 
As can be seen, the decay exponent $\eta$ behaves like 
a typical KT transition; that is, the exponent $\eta$ 
continuously changes with the temperature in the KT phase.  
Since $\eta$ is almost constant for larger $R$ 
in Fig.~\ref{eta}, the exponent at the lower KT temperature 
$T_1$ is estimated as 
$$
   \eta_1 = 0.13(1). 
$$
For smaller $R$ (higher temperature) side, $\eta$  
depends on $R$ due to corrections in Fig.~\ref{eta}.  
Using the fact that the fitted value 
of $b$ in Eq.~(\ref{T_KT}) reflects on the difference from 
the transition point, that is, $L/\xi \propto 1/b$, 
we estimate the exponent $\eta$ at the higher KT temperature $T_2$ 
by extrapolation.  
%%%
The obtained $\eta$ and $b$ are, for example, 
0.310 and 1.76 for $R=0.86$, 0.298 and 2.18 for $R=0.88$, 
and 0.284 and 3.25 for $R=0.90$.  
Plotting the $\eta$'s as a function of $1/b$, 
and extrapolating to $1/b \to 0$, 
we obtain 
$$
   \eta_2 = 0.25(1). 
$$
Of course, other dependences such as $1/b^x$ are possible;  
such an ambiguity is included in the error.  
In Fig.~\ref{eta} we show the value of $R$ which gives $T_2$ 
by the arrow.  The $\eta$ at this $R$ is consistent with 
the estimated value, $0.25(1)$. 
%%%
For the unfrustrated six-state clock model, the exponents 
$\eta_2$ and $\eta_1$ were predicted as 1/4 and 1/9 
respectively \cite{Kosterlitz}, and they were 
confirmed numerically \cite{clock}.  
The present results suggest that the exponents associated 
with the KT transitions are universal even for the frustrated 
model, which the previous work \cite{noh} failed to show.

\subsubsection{Chiral Phase transition}

\begin{figure}
\includegraphics[width=1.0\linewidth]{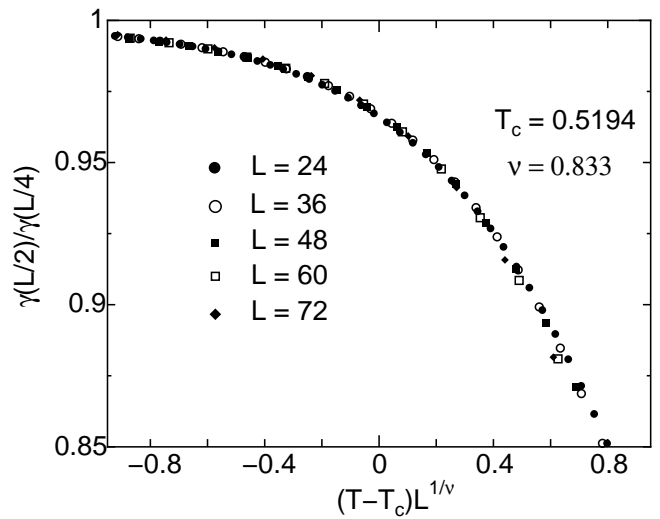}
\caption{FSS plot of the ratio of the chirality 
correlation function, $\gamma(L/2)/\gamma(L/4)$. }
\label{nu1} 
\end{figure}

The temperature dependence of the chiral correlation ratio 
was also plotted in Fig.~\ref{corf}(b).  
The existence of chiral phase transition can be clearly observed. 
In the figure, there is a single crossing point 
which indicates the second-order phase transition; it corresponds to 
the divergent peak in the specific heat plot. 
By using the FSS plot of chirality correlation ratio, 
as shown in Fig.~\ref{nu1}, we can estimate the critical temperature 
and exponent $\nu$ of chiral ordering.  
The estimates are 
$$
  T_c=0.5194(4) \quad {\rm and} \quad \nu=0.83(1).
$$  
Our result exhibiting that the chiral transition occurs 
at slightly higher than $T_2$ of KT transition is consistent 
with most studies in favor of double transition scenario.  
Quite recently, Korshunov \cite{Korshunov} has discussed that 
the phase transition associated with the unbinding 
of vortex pairs takes place at a lower temperature 
than the other phase transition associated 
with proliferation of the Ising-type domain walls.

Our estimate for the exponent $\nu$ is consistent with 
the results by Lee and Lee \cite{lee2} and by Ozeki and Ito \cite{ozeki}, 
but contradicts with the result by Olsson \cite{ols}; 
that is, the critical phenomena are not governed by the Ising 
universality class.  
We have not observed an appreciable size dependence of the estimated $\nu$ 
up to our maximum system size, $L=72$. 
%%%
Olsson \cite{ols} argued that corrections to the scaling 
are important in the fully frustrated XY model, and 
the data are consistent with $\nu = 1$.  Noh {\it et al.} 
also postulated that only for large enough system 
the Ising-like behavior is observed.   
However, using nonequilibrium relaxation study for large enough 
systems up to $L=2000$, Ozeki and Ito \cite{ozeki}
recently obtained the $\nu=0.83(2)$, which suggests that 
the corrections to FSS are not so serious.  
Thus, more careful calculations will be needed 
for the critical phenomena of chiral transition. 
%%%

\section{Concluding Remarks}
In summary, we have investigated the AF six-state clock model 
on the triangular lattice using the Wang-Landau method combined 
with the broad histogram relation.  The model is closely related 
to the 2D fully frustrated XY model.  
We have found that the system possesses two orderings, 
spin ordering and chiral ordering.  The former undergoes 
the KT transition while the latter indicates the second-order 
transition.  We have also observed the lower temperature KT transition 
due to the discrete symmetry of the clock model.  

Our estimates of the higher KT temperature $T_2$ and 
the critical temperature of chiral ordering $T_c$, 
that is, $T_2=0.5154(8)$ and $T_c=0.5194(4)$, 
support the double transition scenario.  
The lower KT temperature is estimated as $T_1=0.496(2)$.  
Two decay exponents of KT transitions are estimated as 
$\eta_2=0.25(1)$ and $\eta_1=0.13(1)$, which suggests that 
the exponents associated with the KT transitions are universal 
even for the frustrated model. 

For the critical phenomena of the chiral transition, our estimate 
of the exponent $\nu$, that is, $\nu=0.83(1)$, suggests that 
the model does not belong to the Ising universality class, 
but more detailed study is still required.

\section*{Acknowledgments}

The authors wish to thank N. Kawashima, H. Otsuka, C. Yamaguchi, 
M. Suzuki, and Y. Ozeki for valuable discussions.  
One of the authors (TS) gratefully acknowledges the fellowship 
provided by the Ministry of Education, Science, Sports and Culture, Japan. 
This work was supported by a Grant-in-Aid for Scientific Research 
from the Japan Society for the Promotion of Science.
The computation of this work has been done using computer facilities 
of Tokyo Metropolitan University and those of 
the Supercomputer Center, Institute for Solid State Physics, 
University of Tokyo.

\end{document}